\documentclass[twocolumn,showpacs,preprintnumbers,amsmath,amssymb,floatfix]{revtex4}
\usepackage{graphicx}% Include figure files
\usepackage{dcolumn}% Align table columns on decimal point
\usepackage{bm}% bold math
\def\beq{\begin{equation}}
\def\eeq{\end{equation}}
\def\be{\begin{eqnarray}}
\def\ee{\end{eqnarray}}

\newcommand{\pom}{\rm I\!P}

\newcommand{\apom}{\alpha_{_{\rm I\!P}}}
\newcommand{\mx}{M_{_X}}

\newcommand{\lsim}{%less than or approx. symbol
 \mathrel{\setbox0=\hbox{$<$}\raise0.6ex\copy0\kern-\wd0
 \lower0.65ex\hbox{$\sim$}}}

\newcommand{\gsim}{%greater than or approx. symbol
 \mathrel{\setbox0=\hbox{$>$}\raise0.6ex\copy0\kern-\wd0
 \lower0.65ex\hbox{$\sim$}}}
 
\newcommand{\T}[1]{{\mathrm{#1}}}

\begin{document}
\title{Mass dependence of nuclear shadowing at small Bjorken-$x$ from diffractive 
scattering}
\author{A. Adeluyi}
\author{G. Fai}
\affiliation{Center for Nuclear Research, Department of Physics \\
Kent State University, Kent, OH 44242, USA}
\date{\today}
\begin{abstract}
We calculate the nuclear shadowing ratio for a wide range of nuclei at small 
Bjorken-$x$ in the framework of Gribov theory. The coherent contribution to the 
(virtual) photon-nucleon cross section is obtained in terms of the diffractive
dissociation cross section. Information on diffraction from FNAL and HERA is
used. Our results are compared to available data from the NMC and E665 
experiments at $x\simeq 10^{-4}$.
\end{abstract}
\maketitle
\vspace{1cm}
%
%%%%%%%%%%%%%%%%%%%%%%%%%%%%%%%%%
% 
\section{Introduction}

Calculations of various cross sections in relativistic nucleus-nucleus 
collision physics in the framework of perturbative QCD (pQCD) require 
nuclear parton distribution functions (nPDFs). The nPDFs are usually 
presented in a parameterized form, as a product of free-nucleon PDFs and a 
``shadowing function''(see e.g. \cite{Eskola:1998df}), which depends on $x$ 
and momentum transfer $Q^{2}$. Experimentally extracted shadowing functions 
are limited by the kinematic range of the data, and fits to these data 
do not provide the shadowing function in the whole requisite ($x$,$Q^{2}$) 
region for all calculations. 

Another approach is to utilize Gribov theory\cite{Gribov:1968gs}, which
relates shadowing to diffraction. For the deuteron, the relationship involves  
the interaction of the diffractively scattered (virtual) photon with only
two nucleons, but for heavy nuclei triple and higher-order scattering
may be important and needs to be included in the formalism. This leads to 
some model dependence. For nuclear collision applications, nPDF uncertainties 
contribute significantly to the overall uncertainties of calculated quantities. 
Thus, an accurate determination of nPDFs is of paramount significance.

It is well-established experimentally that for small values of the Bjorken variable 
($x \lesssim 0.1$), the nuclear structure functions  $F_{2}^{ A}$ are significantly 
reduced compared to $A$ times the free nucleon structure function $F_{2}^{ N}$,
where $A$ denotes the mass number of a specific nucleus.
Equivalently, the virtual photon-nucleus cross section is less than $A$ times 
the one for free nucleons, $\sigma_{\gamma^*  A} < A \,\sigma_{\gamma^*  N}$.
This phenomenon became known as nuclear shadowing (in a strict sense). 
The same behavior is observed for real photons at sufficiently high energies 
($\nu \gtrsim 3\,\rm{GeV}$). Shadowing constitutes the most pronounced nuclear 
effect in lepton-nucleus deep inelastic scattering (DIS). (Note that in a 
generalized sense all modifications of the structure function in the entire 
range of $x$, including regions where $F_{2}^{ A} > A*F_{2}^{ N}$, are sometimes 
referred to as ``shadowing''.)

The shadowing ratio can be defined as $F_{2}^{ A}/(A*F_{2}^{ N})$. Since the 
virtual photon-nucleus cross section is proportional to $F_{2}$, the shadowing 
ratio can be expressed alternatively as $\sigma_{\gamma^* A}/(A*\sigma_{\gamma^* N})$. 
In this work we employ Gribov theory, including the real part of the diffractive 
scattering amplitude, to calculate the shadowing ratio at very small Bjorken-$x$. 
We compare to experimental results from the NMC\cite{Amaudruz:1995tq,Arneodo:1995cs} 
and E665\cite{Adams:1992nf,Adams:1995is} experiments. These experimental data are
all at small $Q^{2}$; we thus use the information from diffractive scattering
of real photons ($Q^{2} = 0$ GeV$^2$).

The paper is organized as follows: in Sec.~\ref{shad} we review the basic 
formalism of Gribov theory as applied to shadowing in the small Bjorken-$x$ 
regime. This is essentially the same treatment as described in 
Ref.~\cite{Frankfurt:2003gx}, where the basic object
is the diffractive structure function. Here the basic quantity is the
differential diffractive cross section. Sec.~\ref{DD} deals with diffractive 
production and the diffractive dissociation
cross section. We present the results of our calculation 
in Sec.~\ref{res}, and conclude in Sec.~\ref{concl}.

%%%%%%%%%%%%%%%%%%%%%%%%%%%%%%%%%%%%%%%%%%%%%%%%%%%%%%%%%%%%%%%%%%%%%%%%%%%%%%%%%%%%%%
%%%%%%%%%%%%%%%%%%%%%%%%%%
\section{Diffraction and Nuclear Shadowing}
\label{shad}

%%%%%%%%%%%%%%%%%%%%%%%%%%%%%%%%
\subsection{Shadowing Ratio}
\label{shadrat}

Several natural length scales may become important in scattering off of a 
nucleus of mass number $A$. Most important for the present discussion is 
the average inter-nucleon separation, $d$, of the order of 2 fm. A high energy 
(virtual) photon scattered from the system can experience nuclear effects in
two distinct ways\cite{Piller:1999wx}:
\begin{enumerate}
\item[i)] 
incoherent scattering from $A$ nucleons with modifications to the structure
functions due to many-body effects in the nuclear medium;
\medskip
\item[ii)] 
coherent scattering processes involving more than one nucleon at a time. 
\end{enumerate}

The latter effects occur when hadronic excitations (or fluctuations) produced 
by the high-energy photon propagate over distances (in the laboratory frame) 
comparable to or larger than the characteristic length scale $d\sim 2$~fm. 
Shadowing can be understood in terms of coherent 
scattering on more than one nucleon. Incoherent scattering occurs primarily 
in the range $0.1 < x < 1$, while strong coherence effects are manifest at
$x<0.1$.

The (virtual) photon-nucleus cross section can be separated into 
a part which accounts for the incoherent scattering 
from individual nucleons, and a correction (shadowing correction) from the 
coherent interaction with several nucleons:
\begin{equation}
\sigma_{\gamma^*{ A}} = 
Z \,\sigma_{\gamma^* { p}} + (A-Z) \,\sigma_{\gamma^* { n}} 
+ \delta \sigma_{\gamma^* { A}}  \,\,\, . 
\end{equation}
The single scattering part is the incoherent sum of 
photon-nucleon cross sections, where $Z$ is the nuclear charge number, 
and $\sigma_{\gamma^* { p}}$ and $\sigma_{\gamma^* { n}}$ are
the photon-proton and photon-neutron cross sections, respectively.
The multiple scattering correction can be viewed as expanded in terms of 
the number of nucleons in the target involved in the coherent scattering 
($n \geq 2$). The leading contribution to nuclear shadowing comes from 
double scattering, since the probability that the propagating hadronic excitation 
coherently interacts with several nucleons decreases with the number of nucleons. 
For the deuteron, only the double-scattering term ($n = 2$) is present.

We define the shadowing ratio as 
\beq
{\mathcal R}^{S}_{A} = \frac{Z \,\sigma_{\gamma^*{p}}+(A-Z)\,\sigma_{\gamma^*{n}} 
+\delta\sigma_{\gamma^*{A}}}{Z \,\sigma_{\gamma^*{p}}+(A-Z)\,\sigma_{\gamma^*{n}}} \,\,\, . 
\eeq 
Thus, the evaluation of the shadowing correction, 
$\delta \sigma_{\gamma^* { A}}$, is central to the
calculation of the shadowing ratio. In the next section we utilize the Gribov
theory in a generalized form to determine $\delta \sigma_{\gamma^* { A}}$.

%%%%%%%%%%%%%%%%%%%%%%%%%%%%%%%%%%%%%%%%
\subsection{Shadowing Correction From Generalized Gribov Theory}
\label{shadcor}

The Gribov theory highlights the deep connection between nuclear shadowing and 
diffraction. The original formulation relates the correction to the cross section 
in hadron-deuteron scattering to diffraction off of a nucleon.
For the deuteron (and other light nuclei where double scattering is dominant), 
the shadowing correction can be evaluated without any ambiguity associated with 
the involvement of more than two target nucleons. For heavier nuclei, one needs 
to go beyond double scattering. Multiple scattering with $n > 2$ nucleons may 
contribute significantly to the shadowing correction (depending on the kinematic 
regime). These contributions are not model-independent; therefore the evaluation 
of the shadowing correction using Gribov formalism depends to some extent on the 
model for the $n > 2$ multiple scattering. 

The original formulation by Gribov neglects the real part of the diffractive scattering
amplitude. Here we take this into account and denote by $\eta$ the 
ratio of the real to imaginary parts of the diffractive scattering amplitude.
In the generalized form incorporating the real part, the shadowing
correction at the level of double scattering is given by 
\begin{eqnarray} \label{eq:ds_A}
\delta \sigma_{\gamma^* { A}}=\frac{A(A-1)}{2A^2} \, 16 \pi \, {\cal R}e 
\Bigg[\frac{(1-i\eta)^2}{1+\eta^2} \nonumber\\
\int d^2 b 
\int^{\infty}_{-\infty} dz_1 
\int^{\infty}_{z_1} dz_2 \int_{4 m_{\pi}^2}^{W^2} dM_{ X}^2  
\left. \frac{d^2\sigma^{{\rm diff}}_{\gamma^{*} { N}}}{dM_{ X}^2 dt}
\right|_{t\approx 0} \nonumber \\ 
\rho_{ A}^{(2)}(\vec b,z_1;\vec b, z_2) \, \, 
\exp{\left\{i\frac{(z_1-z_2)}{\lambda}\right\}} \Bigg] \,. 
\end{eqnarray}
 \begin{figure}
\begin{center} 
\includegraphics[width=4.5cm, height=4.0cm]{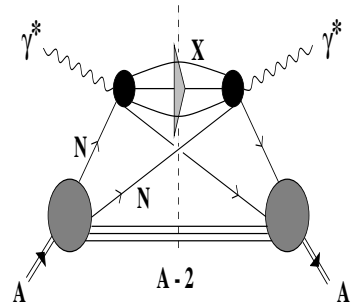}
\end{center}
\caption[...]{Double scattering contribution to nuclear DIS.}
\label{fig:double_scattering}
\end{figure}
Here $\sigma^{{\rm diff}}_{\gamma^{*} { N}}$ is the photon-nucleon 
diffractive cross section and $z$ represents the beam direction.
The coherence length, $\lambda$, is given by 
$\lambda = 2\nu/M_{ X}^2$ for real photons. 
As illustrated in Fig. \ref{fig:double_scattering}, 
a diffractive state with invariant mass $M_{ X}$ is 
produced in the interaction of the photon with  a nucleon 
located at position $(\vec b, z_1)$ in the target. 
The hadronic excitation is assumed to propagate 
at fixed impact parameter $\vec b$ and to interact with a second nucleon at
$z_2$.
The probability to find two nucleons in the target 
at the same impact parameter is described by the two-body density 
$\rho_{ A}^{(2)}(\vec b,z_1;\vec b,z_2)$ normalized as 
$\int d^3 r\,d^3 r'\,\rho_{ A}^{(2)}(\vec r, \vec r\,') = A^2$.
The phase factor, $\exp{\{i[(z_1 - z_2)/\lambda]\}}$ in 
eq.~(\ref{eq:ds_A}) implies that only diffractively  
excited hadrons with a longitudinal propagation length larger than 
the average nucleon-nucleon distance in the target, 
$\lambda > d \simeq 2\,\rm{fm}$, 
can contribute significantly to double scattering.
The limits of integration define the kinematically permitted range 
of diffractive excitations, with their invariant mass $M_{\T X}$ 
above the two-pion production threshold and limited by the 
center-of-mass energy $W=\sqrt{s}$ of the scattering process.

In order to approximate the two-body density 
$\rho_{ A}^{(2)}(\vec b,z_1;\vec b,z_2)$, we note that 
short-range nucleon-nucleon correlations are relevant in nuclei 
when $z_2-z_1$ is comparable to the range of the short-range repulsive part of  
the nucleon-nucleon force, i.e. for distances $\lsim 0.4\,\rm{fm}$. 
However, shadowing is negligible in this case. 
Short-range correlations are therefore not important in the shadowing 
domain, and the target can be considered as an ensemble of 
independent nucleons with  
$\rho_{ A}^{(2)}(\vec r, \vec r\,') 
\approx \rho_{ A}(\vec r) \rho_{ A}(\vec r\,')$, 
where $\rho_{ A}$ is the nuclear one-body density.

With increasing photon energies or decreasing $x$ down to $x\ll 0.1$, 
the longitudinal propagation length of diffractively excited 
hadrons rises and eventually reaches nuclear dimensions. 
Thus, for heavy nuclei interactions of the excited hadronic state with 
several nucleons in the target become important and should be accounted for. 
Following \cite{Frankfurt:2003gx} we introduce an attenuation factor with an 
effective hadron-nucleon cross section, $\sigma_{{\rm eff}}$. The shadowing
correction can thus be written as 
\begin{eqnarray} 
\label{eq:ms_A}
\delta \sigma_{\gamma^* { A}}=\frac{A(A-1)}{2A^2} \, 16 \pi \, {\cal R}e 
\Bigg[\frac{(1-i\eta)^2}{1+\eta^2} \nonumber \\ 
\int d^2 b \int^{\infty}_{-\infty} dz_1
\int^{\infty}_{z_1} dz_2 \int_{4 m_{\pi}^2}^{W^2} dM_{ X}^2 
\left. \frac{d^2\sigma^{{\rm diff}}_{\gamma^{*} { N}}}{dM_{ X}^2 dt}
\right|_{t\approx 0} \nonumber\\
\rho_{ A}^{(2)}(\vec b,z_1;\vec b, z_2) \, 
\exp{\left\{i\frac{(z_1-z_2)}{\lambda}\right\}} \nonumber\\ 
\exp{\left\{-(1/2)(1-i\eta)\sigma_{{\rm eff}} \int_{z_1}^{z_2} dz \rho_A(b,z)\right\}} 
\Bigg] \,. 
\end{eqnarray}
The effective hadron-nucleon cross section, $\sigma_{{\rm eff}}$ in 
eq.~(\ref{eq:ms_A}) is defined as
\begin{equation}
\sigma_{{\rm eff}}=\frac{16 \pi}{\sigma_{\gamma {N}}(1+\eta^2)}
\int_{4 m_{\pi}^2}^{W^2} dM_{ X}^2 \left. \frac{d^2\sigma^{{\rm
diff}}_{\gamma^{*} { N}}}{dM_{ X}^2 dt} \right|_{t\approx 0} \,,
\label{eq:sigeff} 
\end{equation}
where $\sigma_{\gamma N}$ is the photon-nucleon cross section.
The details of this approach and the approximations inherent in the 
definition of $\sigma_{{\rm eff}}$ are treated thoroughly in
\cite{Frankfurt:2003gx}. For vector mesons as the intermediate hadronic 
excitations, we take $\sigma_{VN}$ as $\sigma_{\rm eff}$  in the attenuation 
factor in eq.~(\ref{eq:ms_A}), where $\sigma_{V N}$ is the vector meson-nucleon 
scattering cross section.

When $x \ll 0.1$, it is a good approximation to ignore the phase factor, 
$\exp{i[ (z_1 - z_2)/\lambda]}$, in eq.~(\ref{eq:ms_A}), 
and (using Leibnitz's rule) the integrals over $z_1$ and $z_2$ can be 
carried out explicitly. This leads to a simplified form,
\begin{eqnarray}
\label{eq:s_x}
\delta \sigma_{\gamma^* {A}} =
\frac{2 (1-1/A)\sigma_{\gamma {N}}}{\sigma_{{\rm eff}}} \nonumber\\
{\cal R}e \left(\int d^2 b 
\Bigg[\exp{\left\{-L\,T(b)\right\}}-1+L\,T(b)\Bigg]\right) \,\, , 
\end{eqnarray}
where $L=(A/2) \, (1-i\eta) \, \sigma_{{\rm eff}}$ and 
$T(b)=\int^{\infty}_{-\infty} dz \,\rho_A(b,z)$, the 
usual Glauber thickness function. However, we use the full 
expression, eq.~(\ref{eq:ms_A}), in our calculation.

Eq.~(\ref{eq:ms_A}) gives the shadowing 
correction in terms of $\sigma_{{\rm eff}}$.
This effective cross section involves the differential diffractive
dissociation cross section at small $t$, see eq.~(\ref{eq:sigeff}). 
The treatment of this diffractive cross section is the subject of the next section.

%%%%%%%%%%%%%%%%%%%%%%%%%%%%%%%%%%%%%%%%%%%%%%%%%%%%%%%%%%%%%%%%%%%%%%%%%%%%%%%%%%
%%%%%%%%%%%%%%%%%%%%%%%%%%%%%%%%%%%%%%%%
\section{Diffractive Dissociation}
\label{DD}

%%%%%%%%%%%%%%%%%%%%%%%%%%%%%%%%
\subsection{Diffractive production}
\label{diffprod}

\begin{figure}
\begin{center} 
\includegraphics[width=3cm, height=3.0cm]{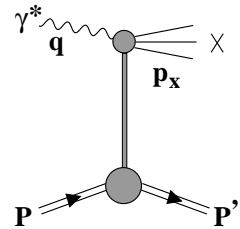}
\end{center}
\caption[...]{Diffractive scattering from a proton.}
\label{fig:singdiff}
\end{figure}

Consider the single diffractive scattering of a (virtual) photon off of a
proton (see Fig. \ref{fig:singdiff}).
The proton does not dissociate, and it remains intact during the process.
The photon, on the other hand, dissociates into a hadronic final state $X$,
which is well separated in rapidity from the proton, 
\begin{equation}
\gamma^{(*)} + { p} \rightarrow  { X} + { p}'. 
\end{equation}
Such diffractive processes are important at small momentum transfer, with 
cross sections which decrease exponentially with the squared four-momentum transfer. 
In general they exhibit a weak energy dependence.

Diffractive dissociation of real photons,
\begin{equation}
\gamma + N \rightarrow  X + N \, ,
\end{equation} 
has been studied in both fixed target and collider experiments. 
Experiments were carried out at Fermi National Laboratory (FNAL)
at average photon-proton center of mass energies of $W \simeq 12.9$ GeV 
and $W \simeq 15.3$ GeV\cite{Chapin:1985mf}. Diffractive states with an 
invariant mass squared of up to $M_{ X}^2 \simeq 18$ GeV$^{2}$ were produced. 
This experiment measured the diffractive dissociation cross section 
differential in both, the invariant mass $M_{ X}$ and the squared four-momentum 
transfer $t$. Experiments at the Hadron-Electron Ring Accelerator
(HERA)\cite{Breitweg:1997eh,Breitweg:1997za,Adloff:1997mi,Aid:1995bz,Derrick:1994dt} 
were carried out at average energies $W \simeq 187$ GeV and $W \simeq 231$ GeV. 
Diffractive states with mass $M_{ X} < 30$ GeV were produced. Unlike the FNAL
experiment, only $d\sigma^{{\rm diff}}_{\gamma^{*} {N}}/dM_{ X}^2$ 
was measured due to poor resolution in $t$. 

As mentioned earlier, the available experimental data on shadowing at small $x$
($x\simeq 10^{-4}$) are all at small $Q^2$ ($Q^2 < 1$ GeV$^2$). At such small
virtualities the photons can be considered quasi-real, and it is thus not a 
bad approximation to regard them as real photons with $Q^2 = 0$ GeV$^2$.
The center-of-mass energies are also low: $W \simeq 15$ GeV for the NMC 
and $W \simeq 25$ GeV for the E665
measurements. These energies are comparable to the photon-proton center of
mass energies at FNAL. For these reasons one can use the information from 
diffractive scattering of real photons at FNAL to calculate the shadowing 
ratio in the kinematic range accessible at NMC and E665.
%%%%%%%%%%%%%%%%%%%%%%%%%%%%%%%%%%%%%%%%%%%%%%%%%%%%

%%%%%%%%%%%%%%%%%%%%%%%%%%%%%%%%%%%%%%%%%%%%%%%%%%%%
\subsection{Diffractive dissociation cross section}
\label{diffxsect}

\begin{figure}
\begin{center} 
\includegraphics[width=6.5cm, height=8.5cm, angle=270]{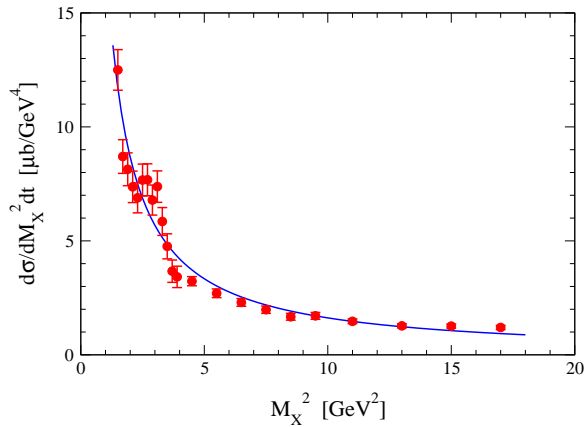}
\end{center}
\caption[...]{(Color Online) Triple-pomeron fit to the FNAL data\cite{Chapin:1985mf},
including the $\rho^{\prime}$ resonances in the continuum, 
as in Ref.~\cite{Piller:1997ny}.}
\label{fig:chaprfit}
\end{figure}

It is natural to divide the diffractive dissociation cross section data as a
function of $M_{X}^2$ into the region of the low-mass vector
mesons ($\rho,\,\omega$, and $\phi$) and a high-mass continuum, with a 
matching point of $M_{X}^2 \simeq 2.25$ GeV$^2$, as in Ref.~\cite{Piller:1997ny}.
However, in this case a triple-pomeron fit to the continuum data (as described
in Sec.~\ref{high-mass}) misses the $\rho^{\prime}$ resonances
as can be seen in Fig.~\ref{fig:chaprfit}.
%%%%%%%%%%%%%%%%%%%%%%%%%%%%%%%%%%%%%%%%%%%%%%%%%%%%

%%%%%%%%%%%%%%%%%%%%%%%%%%%%%%%%%%%%%%%%%%%%%%%%%%%%

The analysis by the H1 collaboration\cite{Adloff:1997mi} divides the 
HERA photoproduction data into effectively three  
intervals in $M_{X}^2$. We follow this strategy in the present paper, 
taking the first interval ($0.16 - 1.58$) GeV$^2$ to contain
the region of the low-mass vector mesons ($\rho,\,\omega$ and $\phi$). 
The second interval ($1.58 - 4.0$ GeV$^2$) covers the $\rho^{\prime}$ 
resonance region. The third interval
($M_{ X}^2 > 4.0$ GeV$^{2}$) is that of the high-mass continuum.
The differential diffractive cross section is thus written as a 
sum over contributions from these three mass intervals,
\begin{eqnarray} 
\label{eq:pDD}
\left. \frac{d \sigma^{D}_{\gamma N}}{d M_X^2 dt} \right|_{t\approx 0} 
= 
\sum_{V=\rho,\omega,\phi}\left. 
\frac{d \sigma^{V}_{\gamma N}}{d M_X^2 dt} 
\right|_{t\approx 0} 
+ 
\sum_{V=\rho^{'}}\left. 
\frac{d \sigma^{V}_{\gamma N}}{d M_X^2 dt} 
\right|_{t\approx 0}  \nonumber \\
+ 
\left.\frac{d \sigma^{cont}_{\gamma N}}{d M_X^2 dt} 
\right|_{t\approx 0}  \,\, . \,\,\,
\end{eqnarray}
In the following, we briefly summarize the various approximations applied
in the three regions.

\subsubsection{Low-mass vector mesons} 
\label{low-mass}

Generalized vector meson dominance (VMD) \cite{Bauer:1977iq} can be utilized to 
describe the contribution of the low-mass vector mesons to the differential 
diffractive cross section, i.e. the first term on the right-hand side 
of eq. (\ref{eq:pDD}):
\begin{equation} 
\label{eq:vect_mes_DD}
\left. \frac{d \sigma^{V}_{\gamma N}}{d M_X^2 dt} \right|_{t\approx 0}
= \frac{e^2}{16\pi}\frac{\Pi^{V}(M_X^2)}{M_X^2} \sigma_{V N}^2.
\end{equation}
with $\Pi^{V}(M_X^2)$ the vector meson part of the photon spectral function 
$\Pi(M_X^2)$, which is given by
\begin{equation}
\Pi(M_X^2) = \frac{1}{12 \pi^2} \frac{\sigma(e^+ e^- \rightarrow hadrons)}
{\sigma(e^+ e^- \rightarrow \mu^+\mu^-)} \,\,\, .
\end{equation}
In eq. (\ref{eq:vect_mes_DD}), $\sigma_{V N}$ is the vector meson-nucleon 
cross section and $e^2/4\pi = 1/137$ is the fine structure constant. 
The $\omega$ and $\phi$ mesons are narrow and thus well 
approximated by delta functions. One can write their contribution 
to the photon spectral function as 
\be
\Pi^{V}(M_X^2) = 
\left(\frac{m_{V}}{g_{V}} \right)^2 
\delta(M_X^2 - m_{V}^2) & ; \,\, & V=\omega,\phi \, ,
\ee
where $m_{V}$ and $g_{V}$, ($V=\omega,\phi$) are the mass and the 
coupling constant of the $\omega$ and $\phi$ mesons, respectively.  

The $\rho$-meson, unlike the $\omega$ and $\phi$ mesons, has a 
large width due to its strong coupling to two-pion states. We have followed
the approach in \cite{Piller:1997ny} and taken this into account through
the $\pi^+\pi^-$ part of the photon spectral function:
\begin{eqnarray} 
\label{eq:Pi_rho}
\Pi^{\rho}\!\left(M_{X}^2\right) = \frac{1}{48 \pi^2} 
\Theta\!\left( M_{X}^2 - 4 m_{\pi}^2 \right) 
\left(1- \frac{4 m_{\pi}^2}{M_{X}^2}\right)^{3/2} \nonumber\\ 
\left| F_{\pi}\left(M_{X}^2\right)\right|^2,
\end{eqnarray}
where $m_{\pi}$ is the mass of the pion
and $M_X=M_{\pi\pi}$ is the invariant mass  
of the $\pi^+\pi^-$ pair. The pion form factor, $F_{\pi}$ is taken from 
Ref. \cite{Klingl:1996by}. A full discussion is given in \cite{Piller:1997ny}. 
We compared the result from the delta function approximation to this more exact
calculation and found that taking into account the width of the
$\rho$-meson increases the differential diffractive cross section by 10\%.

The vector meson-nucleon cross section in eq. (\ref{eq:vect_mes_DD})
has an energy dependence of the form
\begin{equation} \label{eq:vnenergy} 
\sigma_{VN}\sim W^{2(\apom(0) -1)} = W^{2\epsilon}
\end{equation}
where $\apom(t=0) = 1+ \epsilon$ is the soft pomeron 
intercept\cite{Donnachie:1992ny}. 

\subsubsection{Region of the $\rho^{\prime}$ resonances}
\label{rhop-res}

The $\rho^{\prime}$ resonance region contains the $\rho(1450)$ and
$\rho(1700)$ mesons. These resonances were formerly
classified as the $\rho(1600)$ \cite{Hagiwara:2002fs}. 
In Ref.~\cite{Piller:1997ny} the $\rho^{\prime}$ resonance region is 
lumped together with the high-mass continuum. As can be seen 
in Fig.~\ref{fig:chaprfit}, the FNAL data show an enhancement in this region. 
We treat this enhancement in terms of an average $\rho^{\prime}$ 
resonance, corresponding to the earlier classification of $\rho(1600)$, as done 
in Ref.~\cite{Chapin:1985mf}. We use the available information on the  
$\rho(1600)$ from Ref. \cite{Bauer:1977iq} in a VMD-type calculation to
evaluate the contribution from this region. The average $\rho^{\prime}$
resonance should have a finite width, but encouraged by the fact that
a delta function in the case of the $\rho$ gives a good
approximation to the full-width result, we employ a narrow resonance
approximation for the $\rho(1600)$. Thus, for the second term of
(\ref{eq:pDD}) we have
\begin{equation} 
\label{eq:vect_mes_D2}
\left. \frac{d \sigma^{V}_{\gamma N}}{d M_X^2 dt} \right|_{t\approx 0}
= \frac{e^2}{16\pi}\frac{\Pi^{V}(M_X^2)}{M_X^2} \sigma_{V N}^2.
\end{equation}
with
\begin{equation}
\Pi^{V}(M_X^2) = 
\left(\frac{m_{V}}{g_{V}} \right)^2 
\delta(M_X^2 - m_{V}^2)
\end{equation}
and $V=\rho(1600)$.

\subsubsection{High-mass continuum} 
\label{high-mass}

A full treatment of both the FNAL data and the HERA data in this region has
been carried out by the H1 Collaboration in Ref. \cite{Adloff:1997mi}, using 
the triple-Regge model of photon dissociation. 
For simplicity we consider only the triple-pomeron term, ignoring the
sub-leading reggeons and interference between the pomeron and reggeons. In this
case the differential dissociation cross section can be written as
\begin{eqnarray}
\label{eq:t_pom}
  \frac{ {\rm d^2} \sigma}{{\rm d}\mx^2 \, {\rm d}t} =
  \frac{G_{\pom \pom \pom}(0)}{s_{_0}^{\apom(0) - 1}} \
  \left( W^2 \right)^{2 \apom(0) - 2} \  \nonumber \\
  \left( \frac{1}{\mx^2} \right)^{\apom(0)} \
  e^{B (W^2, \mx^2) \: t} \ ,
\end{eqnarray}
where $s_{0}$ is the hadronic mass scale
and $B(W^2, \mx^2) = 2b_{p \pom}+b_{\pom \pom \pom}+2
\apom^{\prime} \ln W^2 / \mx^2$. 
Here $b_{p \pom}$ is the proton-pomeron
slope parameter, $b_{\pom \pom \pom}$ the triple pomeron slope parameter, and
$\apom^{\prime}$ is the slope of the pomeron trajectory. We use the values in 
\cite{Adloff:1997mi} for these parameters, with $s_{0} = 1$ GeV$^2$.

\begin{figure}
\begin{center} 
\includegraphics[width=6.5cm, height=8.5cm, angle=270]{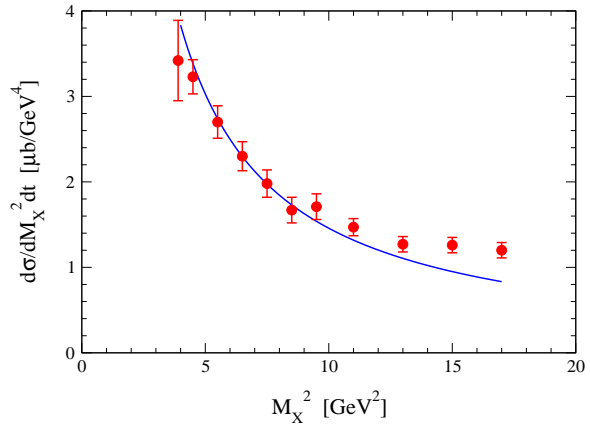}
\end{center}
\caption[...]{(Color Online) Triple pomeron fit to the FNAL\cite{Chapin:1985mf} data
in the nonresonant continuum ($M_X^2 > 4$ GeV$^2$).}
\label{fig:chapfit}
\end{figure}

In order to have a consistent treatment of the different contributions, we take 
$\apom(0) = 1.068 \pm 0.0492$ from \cite{Adloff:1997mi}. This agrees within error
with the soft pomeron intercept in Ref. \cite{Donnachie:1992ny} 
($\apom(0) \simeq 1.081$). Thus the only free parameter is
the triple-pomeron coupling $G_{\pom \pom \pom}(0)$. We fit the FNAL data at 
the average energy $W = 14.3$ GeV and $t = -0.05$ GeV to eq.~(\ref{eq:t_pom}).
The fit gives $G_{\pom \pom \pom}(0) = 11.4$ $\mu$b/GeV$^{2(1+\epsilon)}$. 
The quality of our fit is shown in Fig.~\ref{fig:chapfit}.
%%%%%%%%%%%%%%%%%%%%%%%%%%%%%%%%%%%%%%%%%%%%%%%%%%%%%%%%%%%%%%%
%%%%%%%%%%%%%%%%%%%%%%%%%%%%%%%%%%%%%%%%%%%%%%%%%%%%%%%%%%%%%%%

%%%%%%%%%%%%%%%%%%%%%%%%%%%%%%%%%%%%%%%%%%%%%%%%%%%%%%%%%%%%%%%
%%%%%%%%%%%%%%%%%%%%%%%%%%%%%%%%%%%%%%
\section{Results}
\label{res}

The treatment outlined in the last two sections is now applied to 
calculate the shadowing correction, and hence the shadowing ratio. 
The basic equation is eq.~(\ref{eq:s_x}), 
which involves the ratio of the real to imaginary
amplitudes $\eta$, the photon-nucleon cross section $\sigma_{\gamma {N}}$, 
the nuclear density $\rho_{A}$, and
the effective cross section $\sigma_{{\rm eff}}$.

We use the energy-independent $\eta$'s for the vector mesons from
Ref.~\cite{Bauer:1977iq}. For both $\rho$ and $\omega$ mesons, $\eta$ takes 
values between $0$ and $-0.3$. Here we take $\eta_{\rho} = \eta_{\omega} = -0.2$ 
in accordance with Ref.~\cite{Bauer:1977iq}. The results of our calculation are not
very sensitive to the precise values of $\eta_{\rho(\omega)}$. For the $\phi$ meson,
we take $\eta_{\phi} = 0.13$ \cite{Frankfurt:1998vx}. For lack of information, 
we take $\eta_{\rho(1600)} = 0$.
For the high-mass continuum, we follow Ref.~\cite{Frankfurt:2003gx} and define 
$\eta_{cont}$ as
\begin {equation}
\eta_{cont} = \frac{\pi}{2}\big(\apom (0)-1\big) \,,
\end{equation}
using the result of Gribov and Migdal \cite{Gribov:1968uy}.

The small difference between the photon-proton cross section $\sigma_{\gamma {p}}$ 
and the photon-neutron cross section $\sigma_{\gamma {n}}$ is neglected in this study. 
We use the Donnachie-Landshoff parameterization of 
$\sigma_{\gamma {p}}$\cite{Donnachie:1992ny} as the generic photon-nucleon 
cross section $\sigma_{\gamma {N}}$.
For the nuclear densities three-parameter Fermi ($3pF$) distributions are applied:
\begin{equation}
\rho (r) = \rho_{0} \frac{1 + \omega(r/R_{A})^2}{1 + e^{(r-R_{A})/d}} \,\, ,
\end{equation}
with the parameter values taken from Ref.~\cite{DeJager:1974dg}. For mass
numbers $A \lesssim 20$ a harmonic oscillator (HO) density distribution may be
more appropriate than the $3pF$ distribution. For uniformity, we use
the $3pF$ distributions for the whole mass range in light of the fact that
uncertainties associated with other parameters are at least comparable.
\begin{figure}[h]
\begin{center} 
\includegraphics[width=6.5cm, height=8.5cm, angle=270]{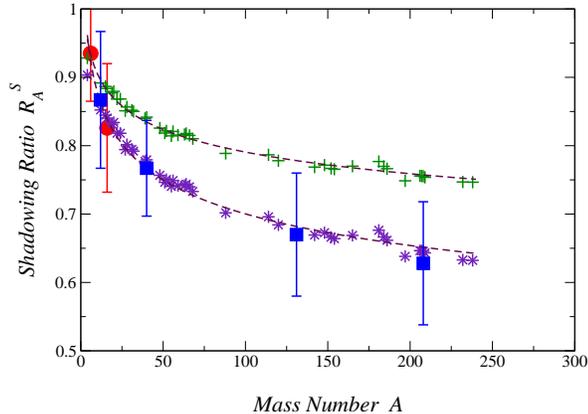}
\end{center}
\caption[...]{(Color Online) Shadowing ratio calculated at $W=15$~GeV (crosses) and
$W=25$~GeV (stars). Data are from the NMC (shaded circles) and E665 (shaded boxes) 
collaborations.
The NMC point corresponding to $^{12}$C is displaced slightly for better visibility.
The dashed lines are energy-dependent fits according to (\ref{shadfit}) as described 
in the text.}
\label{fig:tshad}
\end{figure}

We carried out calculations at $W = 15$ and $25$ GeV, the approximate 
energies of the NMC and E655 experiments, respectively. The results are 
displayed in  Fig.~\ref{fig:tshad}, together with the experimental data points. 
At very small $x$ ($x\simeq10^{-4}$), NMC has two data points, corresponding 
to $^6$Li and $^{12}$C. The E665 experiment has four data points: 
$^{12}$C, $^{40}$Ca, $^{131}$Xe, and $^{208}$Pb. In view of the large error bars 
of the data, the calculation can be said to describe the experimental information 
in the entire mass range at E665 energies, and also for the two available data points 
at NMC energies. For small $A$ the shadowing ratio decreases rapidly with
$A$, while for large $A$ the decrease is more gradual. The calculated result
also falls steeply with increasing $W$.

The dashed lines in Fig.~\ref{fig:tshad} represent a two-parameter
fit of the standard form
\beq
\label{shadfit}
{\mathcal R}^{S}_{A} = \beta_{0}A^{\beta_{1}-1}
\eeq
to the calculated results, with energy-dependent parameters $\beta_{0}$ and 
$\beta_{1}$. In order to determine the energy dependence of the fit parameters
we calculated the shadowing ratio for $W$ in the range $10 \leq W \leq 30$ GeV.
The energy dependence of the fit coefficients can be described as
\beq
%\beta_{0} = 0.88489 + 0.015522W - 0.0002176W^{2}
\beta_{0} = 0.720 + 0.118\ln(W)
\eeq
and
\beq
%\beta_{1} = 1.0467 - 0.0099313W + 0.00014083W^{2}
\beta_{1} = 1.143 - 0.075\ln(W) \,\,   ,
\eeq  
where $W$ is the center-of-mass energy in GeV.
 
The fit does very well for the entire mass range for 
low $W$ and deviates from the calculated result at large $A$ as $W$ 
increases. This seems to suggest that the mass dependence of the calculated 
shadowing ratio at large $A$ and increasing $W$ is not as simple as in 
eq.~(\ref{shadfit}). A five-parameter fourth-degree polynomial with 
energy-dependent coefficients gives a good fit for the entire mass range and 
at all energies considered. However, we prefer the simple physical 
picture of the two-parameter fit, which is adequate considering the
experimental and theoretical error bars. 

The uncertainties of our calculation are mostly related to the various 
parameterizations of the diffractive dissociation cross section. The
delta function parameterization for the $\omega$ and $\phi$ mesons 
should be satisfactory, and the width of the $\rho$ meson has been 
taken into account. Refinements of the spectral function (\ref{eq:Pi_rho})
are possible, and improvements of the treatment of the $\rho^{\prime}$ resonance 
region is also left for future work. The uncertainties in the continuum are 
associated with the neglect of sub-leading reggeons and of interference terms.
Furthermore, the use of an effective scattering cross
section to account for multiple scattering is an approximation,
as is using real-photon information ($Q^{2} = 0$) at small, but 
non-vanishing $Q^{2}$. We estimate
the overall uncertainty of our calculated results to be in the 20\% range. 

%%%%%%%%%%%%%%%%%%%%%%%%%%%%%%%%%%%%%%%
\section{Conclusion}
\label{concl}

We have calculated the shadowing ratio at very small Bjorken-$x$ for nuclei 
in the range $3 < A < 239$ using Gribov theory. We included the effect of the 
real part of the diffractive scattering amplitude. The photon diffractive 
dissociation cross section, which serves as an input to our calculation, was
parameterized as a function of the invariant mass of the diffractively produced
hadronic excitation using vector meson dominance and Regge theory in three mass intervals:
low-mass vector mesons, $\rho^{\prime}$ resonances, and continuum. The 
parameters needed are taken from earlier studies that fit experimental data.

It is found that the calculated shadowing ratio decreases with mass number
first rapidly and then more slowly. To be able to compare to NMC/E665 data, 
the calculations are all at low center-of-mass energy, and the decrease with 
$A$ is stronger as the center-of-mass energy increases. We find that Gribov theory
gives a reasonable estimate of the mass dependence of nuclear shadowing at 
small Bjorken-$x$.  
%
%%%%%%%%%%%%%%%%%%%%%%%%%%%%%%%%%%%%%%%%
%

\end{document}